\documentclass[aps, preprint, superscriptaddress]{revtex4}
\usepackage{graphicx, epsfig}
\usepackage{epstopdf}

\DeclareGraphicsExtensions{.  jpg, .  pdf,  .  mps,  . png,  . eps,  .  ps,  .  EPS}
\usepackage{amsmath}
\usepackage{color}
\begin{document}
\newcommand{\avg}[1]{\langle{#1}\rangle}
\newcommand{\ket}[1]{\left |{#1}\right \rangle}
\newcommand{\beq}{\begin{equation}}
\newcommand{\eneq}{\end{equation}}
\newcommand{\beqnn}{\begin{equation*}}
\newcommand{\eneqnn}{\end{equation*}}
\newcommand{\beqy}{\begin{eqnarray}}
\newcommand{\eneqy}{\end{eqnarray}}
\newcommand{\beqynn}{\begin{eqnarray*}}
\newcommand{\eneqynn}{\end{eqnarray*}}
\newcommand{\half}{\mbox{$\textstyle \frac{1}{2}$}}
\newcommand{\proj}[1]{\ket{#1}\bra{#1}}
\newcommand{\av}[1]{\langle #1\rangle}
\newcommand{\braket}[2]{\langle #1 | #2\rangle}
\newcommand{\bra}[1]{\langle #1 | }
\newcommand{\Avg}[1]{\left\langle{#1}\right\rangle}
\newcommand{\inprod}[2]{\braket{#1}{#2}}
\newcommand{\upket}{\ket{\uparrow}}
\newcommand{\downket}{\ket{\downarrow}}
\newcommand{\Tr}{\mathrm{Tr}}
\newcommand{\hcontrol}{*!<0em, . 025em>-=-{\Diamond}}
\newcommand{\hctrl}[1]{\hcontrol \qwx[#1] \qw}
\newenvironment{proof}[1][Proof]{\noindent\textbf{#1. } }{\ \rule{0. 5em}{0. 5em}}
\newtheorem{mytheorem}{Theorem}
\newtheorem{mylemma}{Lemma}
\newtheorem{mycorollary}{Corollary}
\newtheorem{myproposition}{Proposition}
\newcommand{\vp}{\vec{p}}
\newcommand{\Or}{\mathcal{O}}
\newcommand{\so}[1]{{\ignore{#1}}}

\newcommand{\red}[1]{\textcolor{red}{#1}}
\newcommand{\blue}[1]{\textcolor{blue}{#1}}

\newcommand{\bea}{\begin{eqnarray}}
\newcommand{\eea}{\end{eqnarray}}
\newcommand{\gt}{\tilde{g}}
\newcommand{\mt}{\tilde{\mu}}
\newcommand{\et}{\tilde{\varepsilon}}
\newcommand{\ct}{\tilde{C}}
\newcommand{\bt}{\tilde{\beta}}

\title{  Complex Quantum Network Manifolds in Dimension $d>2$  \\are Scale-Free}

\author{ Ginestra Bianconi}
\affiliation{School of Mathematical Sciences,  Queen Mary University of London,  London E1 4NS,  United Kingdom}
\author{Christoph Rahmede}
\affiliation{Institute for Theoretical Physics,  \\Karlsruhe Institute of Technology,  76128 Karlsruhe,   Germany}

\begin{abstract}
{\bf
In quantum gravity, several  approaches have been  proposed until now  for the quantum description of discrete geometries. These theoretical frameworks include loop quantum gravity,  causal dynamical triangulations,  causal sets,  quantum graphity,  and energetic spin networks. Most of these approaches describe discrete spaces as   homogeneous   network manifolds. Here we define Complex Quantum  Network Manifolds (CQNM) describing the evolution of quantum network states,   and constructed from growing simplicial complexes of dimension $d$. We show that in $d=2$ CQNM are homogeneous networks while for $d>2$ they are scale-free   i.e.  they are characterized by large inhomogeneities of degrees like most complex networks. From the self-organized   evolution of CQNM quantum statistics 
emerge spontaneously.  Here we define  the generalized degrees associated with the  $\delta$-faces of  the  $d$-dimensional CQNMs,  and we show that the statistics of these generalized degrees can either follow Fermi-Dirac,  Boltzmann or Bose-Einstein distributions depending on the dimension of the $\delta$-faces. }
\end{abstract}
\maketitle


Several theoretical approaches have been proposed in quantum gravity for the description  and characterization of quantum discrete spaces including  loop quantum gravity \cite{Smolin_Rovelli, Smolin_Rovelli2, Rovelli},  
causal dynamical triangulations \cite{CDT1, CDT2},  causal sets \cite{Rideout,Eichhorn:2013ova},  quantum graphity  \cite{graphity_rg, graphity, graphity_Hamma},   energetic spin networks \cite{Cortes1, Cortes2}, and diffusion processes on such quantum geometries  \cite{Calcagni:2013vsa}.
In most of these approaches,  the discrete spaces are network manifolds with  homogeneous degree distribution and do not have common features with complex networks describing complex systems such as the brain or the biological networks in the cell.     
Nevertheless it has been discussed  \cite{Lifecosmos} that a consistent theory of  quantum cosmology could also  be a theory of self-organization \cite{Bak_Sneppen, jensen},   sharing some of its dynamical properties with  complex systems and biological evolution. 

In the last decades,  the field of  network theory \cite{BA_review, Newman_book, Doro_book, Boccaletti2006,Caldarelli_book} has made significant advances in the  understanding of the underlying network 
topology of complex systems as diverse as the biological networks in the cell,  the brain networks,   or the Internet.  Therefore an increasing interest is addressed to  the study of quantum gravity 
from the information theory and  complex network perspective \cite{Cosmology,Trugenberger:2015xma}. 
 
In network theory it  has been found that scale-free networks  \cite{BA} characterizing highly inhomogeneous network structures  are  ubiquitous and  characterize biological,  technological and social 
systems \cite{BA_review, Newman_book, Doro_book, Boccaletti2006}.  Scale-free networks have finite average degree but infinite fluctuation of the degree distribution and in these structures nodes 
(also called "hubs") with a number of connections much bigger than the average degree  emerge.  Scale-free networks  are known to  be robust to random perturbation  and 
there is a significant interplay between structure and dynamics, since critical phenomena such as in the Ising model,
synchronization or epidemic spreading change their phase diagram when defined on them \cite{doro_crit, Dynamics}. 
 
Interestingly,  it has been shown that such networks,  when they are evolving by a dynamics inspired by biological evolution,   can be described by the Bose-Einstein statistics,  and they might undergo 
a Bose-Einstein condensation in which a  node is linked to a finite fraction of all the nodes of the network \cite{Bose}.  Similarly evolving Cayley trees  have been shown to follow a Fermi-Dirac 
distribution  \cite{Fermi, Complex}. 
 
Recently,  in the field of complex networks  increasing attention is devoted to the characterization of the geometry of complex networks 
\cite{Aste, Kleinberg, Boguna_navigability, Boguna_hyperbolic, Saniee, Mason, Aste2,Vaccarino1, Vaccarino2,Caldarelli_geometry}.  In this context,  special attention has been addressed to simplicial 
complexes \cite{Emergent, Q, Farber,  Kahle, Dima_simplex},  i.e.  structures formed by gluing together simplices  such as triangles,  tetrahedra etc. 

Here we focus our attention on Complex Quantum Network Manifolds (CQNMs) of dimension $d$  constructed by gluing together simplices of dimension $d$.  The CQNMs grow according to a  non-equilibrium 
dynamics  determined by the energies associated to its nodes, and have an emergent  geometry, i.e. the geometry of the CQNM is not imposed a priori  on the network manifold, but it is determined by its stochastic dynamics.
  Following a similar procedure as used in several other manuscripts \cite{graphity_rg, graphity, graphity_Hamma,Q},  one can show that the  
CQNMs characterize the time evolution of the quantum network states.  In particular,  each network evolution can be considered as a possible path over which the path integral characterizing the 
quantum network states can be calculated.  
 Here we show that in $d=2$ CQNMs are homogeneous and have an exponential degree distribution while    the CQNMs are always scale-free for $d>2$. Therefore for $d=2$ the degree distribution of the CQNM has bounded fluctuations and is homogeneous while for $d=2$ the CQNM has unbounded fluctuations in the degree distribution and its structure is dominated by hub nodes.
 Moreover,  in  CQNM quantum statistics emerges spontaneously from the network dynamics. 
In fact, here we define the generalized degrees of the  $\delta$-faces forming the manifold and we show that the average of the generalized degrees of the $\delta$-faces  with energy $\epsilon$ 
follows different statistics (Fermi-Dirac,  Boltzmann or Bose-Einstein statistics) depending on the dimensionality $\delta$ of the faces and  on the dimensionality $d$ of the CQNM. 
For example in $d=2$ the average of the generalized degree of the links follows a Fermi-Dirac distribution and the average of the generalized degrees of the  nodes follows a Boltzmann distribution. 
In $d=3$ the faces of the tetrahedra,  the links and the nodes have an average of their generalized degree that follows respectively the Fermi-Dirac distribution,   the Boltzmann distribution and the Bose-Einstein distribution.

Consider a $d$-dimensional simplicial complex formed by gluing together simplices of dimension $d$,  i.e.  a triangle for $d=2$,  a tetrahedron for $d=3$ etc.  
A necessary requirement for obtaining a discretization of a manifold  is that 
each simplex of dimension $d$ can be glued to another simplex  only in such a way that the  $(d-1)$-faces formed by $(d-1)-$dimensional simplices (links in $d=2$,  triangles in $d=3$,  etc.) 
belong at most to two simplices of dimension $d$.

Here we  indicate with ${\cal S}_{d, \delta}$ the set of all $\delta$-faces  belonging to the $d$-dimensional manifold with $\delta<d$. 
If  a $(d-1)$-face $\alpha$   belongs to two simplices of dimension $d$ we will say that  it is "saturated" and we indicate this by an associated variable $\xi_\alpha$ with value $\xi_{\alpha}=0$; if it belongs to  only one simplicial complex of dimension $d$  we will say that it is "unsaturated" and we will indicate this by setting $\xi_{\alpha}=1$. 

The CQNM is evolving according to a non-equilibrium dynamics described in the following. 

\begin{figure}
\begin{center}
{\includegraphics[width=1\columnwidth]{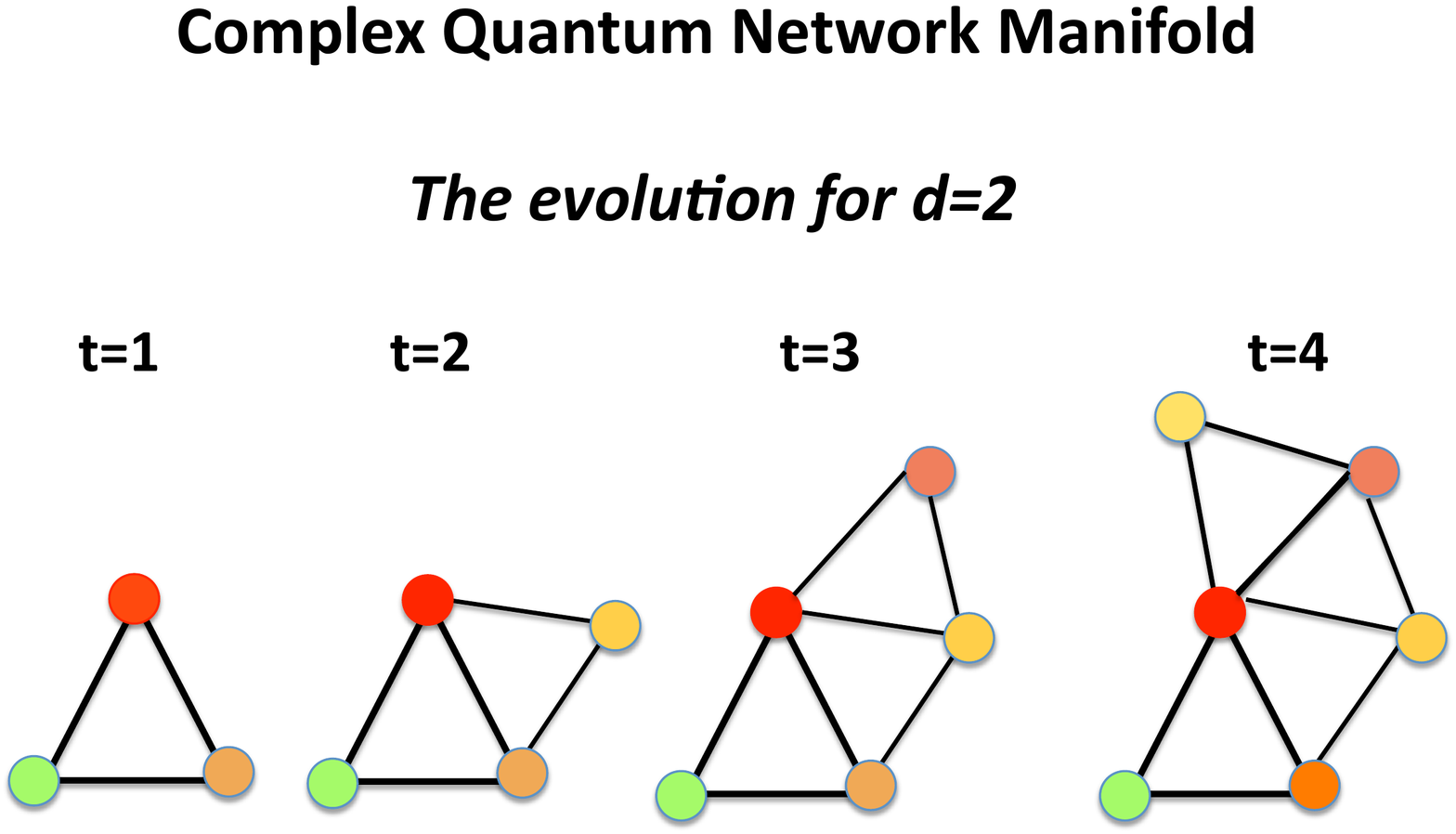}}
\end{center}
\caption{Few steps of a possible evolution of the CQNM for $d=2$.  The nodes have different energies represented as different colours of the nodes.  A link can be saturated (if two triangles are adjacent to it) or unsaturated (if only one triangle is incident to each).  Starting from a single triangle at time $t=1$,  the CQNM evolves through the addition of new triangles to unsaturated links.  }
\label{fig1}
\end{figure}

\begin{figure}
\begin{center}
{\includegraphics[width=1\columnwidth]{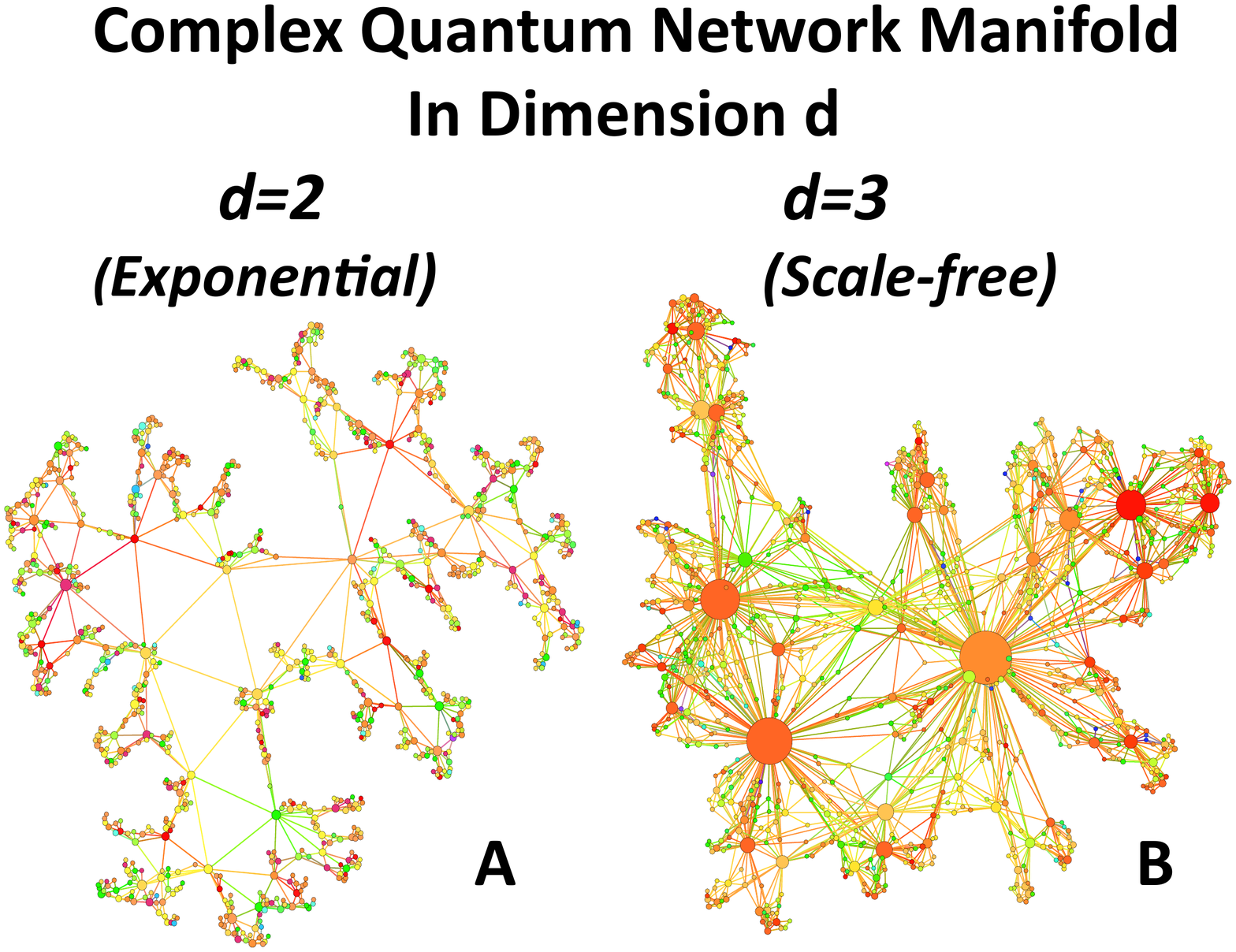}}
\end{center}
\caption{Visualization of CQNM with  $d=2$ (panel A) and  $d=3$ (panel B).  The colours of the nodes indicate their energy $\{\epsilon_i\}$ while their size indicates their degree $K_d(i)$.  In $d=2$ the degree distribution of the CQNMs is a convolution of exponentials,  in $d=3$ the CQNMs are scale-free and the presence of hubs is clearly observable from this visualization.  The data shown are for CQNM with $N=10^3$ nodes,  $\beta=0. 2$ and Poisson distribution $g(\epsilon)$ with average $z=5$. }
\label{fig2}
\end{figure}

To each node $i=1, 2\ldots,  N$ an {\em energy } of the node $\epsilon_i$ is assigned from a distribution $g(\epsilon)$.  The energy of the node is quenched and does not change during the evolution of the network. 
To every  $\delta$-face  $\alpha\in {\cal S}_{d, \delta}$ we associate an {\em energy} $\epsilon_{\alpha}$ given by the sum of the energy of the nodes that belong to the face  $\alpha$,
\bea
\epsilon_{\alpha}=\sum_{i \subset \alpha}\epsilon_i. 
\eea
At time $t=1$ the CQNM is formed by a single $d$-dimensional simplex. 
At  each time $t>1$ we add a simplex of dimension $d$ to an unsaturated $(d-1)$-face $\alpha\in{\cal S}_{d, d-1}$ of dimension $d-1$.  We choose this simplex with   probability  $\Pi_{\alpha}$ given by  
\bea
\Pi_{\alpha}&=&\frac{1}{Z}e^{-\beta \epsilon_{\alpha}}\xi_{\alpha}, 
\label{P1}
\eea
where $\beta$ is a parameter of the model called {\em inverse temperature} and $Z$ is a normalization sum given by 
\bea
Z=\sum_{\alpha\in {\cal S}_{d, d-1}}e^{-\beta \epsilon_{\alpha}}\xi_{\alpha}. 
\eea
Having chosen the $(d-1)$-face $\alpha$,  we glue to it a new $d$-dimensional simplex containing  all the nodes of the $(d-1)-$face $\alpha$  plus the new node $i$.  It follows that the new node $i$ is linked to each node $j$ belonging to $\alpha$.  \\
In Figure $\ref{fig1}$ we show few steps of the evolution of a CQNM for the case $d=2$,  while in Figure $\ref{fig2}$ we show examples of CQNM in $d=2$ and in $d=3$ for different values of $\beta$. \\
From the definition of the non-equilibrium dynamics described above, it is immediate to show that the network structure constructed by this non-equilibrium dynamics is connected and is a discrete manifold.\\
Since at time $t$ the number of nodes of the network manifold is $N=t+d$, the  evolution of the network manifold is fully determined  by the  sequence  $\{\epsilon_i\}_{i\leq t+d}$, and the sequence  $\{\alpha_{t'}\}_{t' \leq \ldots t}$, where  $\epsilon_i$, for $i\leq d+1$ indicates the energy of an initial node, while for $i=t'+d$ with $t'>1$ it indicates the energy of the node added at time $t'$, and   where $\alpha_{t'}$ indicates  the $(d-1)$-face to  which the new $d$-dimensional simplex is added at time $t'>1$. \\
The dynamics described above  is inspired by biological evolutionary dynamics  and is related to  self-organized critical models.  In fact the case $\beta\to \infty$ is dictated by an extremal dynamics that can be related to invasion percolation \cite{IP, Fermi},  while the case $\beta=0$ can be identified as an Eden model \cite{Eden} on a $d$-dimensional simplicial complex.  \\

Here we call these network manifolds  Complex Quantum Network Manifolds because using similar arguments already developed in \cite{graphity_rg, graphity, graphity_Hamma, Q} it can be shown that they  describe the evolution of Quantum Network States (see  Supplementary Information for details). 
The quantum network state  is an element of an Hilbert space ${\cal H}_{tot}$ associated to a simplicial complex of $N$ nodes formed by gluing $d$-dimensional simplices  (see  Supplementary Information for details). 
The quantum network state $\ket{\psi(t)}$ evolves through a Markovian non-equilibrium dynamics determined by the energies $\{\epsilon_i\}$ of the nodes.  
The quantity ${\cal Z}(t)$  enforcing the normalization of the quantum network state $\langle \psi(t)|\psi(t)\rangle=1$ 
 can be interpreted as a path integral over CQNM evolutions determined by the sequences $\{\epsilon_{i}\}_{i\leq t+d}, $ and $\{\alpha_{t'}\}_{t'\leq t}$. 
 In fact we have 
\bea
{\cal Z}(t)=\sum_{\{\epsilon_i\}_{i\leq  t+d}}\sum_{\{ \alpha_{t'}\}_{t'\leq t}}W(\{\epsilon_{i}\}_{i\leq t+d}, \{\alpha_{t'}\}_{t'\leq t}),
\eea
where the explicit expression of $W(\{\epsilon_{i}\}_{i\leq t+d}, \{\alpha_{t'}\}_{t'\leq t})$ is given in the Supplementary Information.
Moreover,   ${\cal Z}(t)$ can be interpreted as the partition function of the statistical mechanics problem over the CQNM temporal evolutions. 
If we identify the sequences  $\{\epsilon_{i}\}_{i\leq t+d}, $ and $\{\alpha_{t'}\}_{t'\leq t}$,  determining ${\cal Z}(t)$ with the sequences  indicating the temporal evolution of the CQNM we have that the probability $P(\{\epsilon_{i}\}_{i\leq t+d}, \{\alpha_{t'}\}_{t'\leq t})$ of a given CQNM evolution is given by 
\bea
P(\{\epsilon_{i}\}_{i\leq t+d}, \{\alpha_{t'}\}_{t'\leq t})=\frac{W(\{\epsilon_{i}\}_{i\leq t+d}, \{\alpha_{t'}\}_{t'\leq t})}{{\cal Z}(t)}.
\eea
Therefore each classical evolution of the CQNM up to time $t$ corresponds to one of the paths defining the evolution of the quantum network state up to time $t$. \\
A set of  important structural properties of the CQNM are the generalized degrees $k_{d, \delta}$ of its $\delta$-faces. 
Given a CQNM of dimension $d$,  the generalized degree $k_{d, \delta}(\alpha) $ of a given $\delta$-face $\alpha$,  (i.e.  $\alpha\in {\cal S}_{d, \delta}$) is defined as the number of $d$-dimensional simplices  incident to it. 
For example,  in a CQNM of dimension $d=2$,  the generalized degree $k_{2, 1}(\alpha)$ is the number of triangles incident to a link $\alpha$ while the generalized degree $k_{2, 0}(\alpha)$ indicates the number of triangles incident to a node $\alpha$. 
Similarly in a CQNM of dimension $d=3$,  the generalized degrees $k_{3, 2}, k_{3, 1}$ and $k_{3, 0}$ indicate  the number of tetrahedra incident respectively to a triangular face,  a link or a node. 
If from a CQNM of dimension $d$ one extracts the underlying network,  the degree $K_d(i)$ of node $i$ is given by the generalized degree $k_{d, 0}(i)$ of the same node $i$ plus $d-1$,  i.e. 
\bea
K_d(i)=k_{d, 0}(i)+d-1. 
\label{KD}
\eea
We indicate with 
$P_{d, \delta}(k)$ the distribution of generalized degrees $k_{d, \delta}=k$. 
It follows that the degree distribution of the network 
$P_d(K)$ constructed from the  $d$-dimensional CQNM is  given by 
\bea
P_d(K)=P_{d, 0}(K-d+1). 
\label{Pk}
\eea
Let us consider the generalized degree distribution of CQNM in  the case $\beta=0$.  In this case the new $d$-dimensional simplex can be added with equal probability to each unsaturated $(d-1)$-face  
of the CQNM. 
Here we show that as long as the dimension $d$ is greater than two,  i.e.  $d>2$, the CQNM is a scale-free network. 
In fact each $\delta$-face,  with  $\delta<d-1$,  which has generalized  degree $k_{d, \delta}(\alpha)=k$,  is incident to 
\bea
2+(d-\delta-2)k
\eea
unsaturated $(d-1)$-faces.  Therefore   the probability 
$\pi_{d, \delta}(\alpha)$ to attach a new $d$-dimensional simplex to a $\delta$-face $\alpha$ with generalized degree $k_{d, \delta}(\alpha)$ and with $\delta<d-1$,  is given by 
\bea
\pi_{d, \delta}(\alpha)=\frac{2+(d-\delta-2)k_{d,\delta}(\alpha)}{\sum_{\alpha'\in {\cal S}_{d,\delta}}[2+(d-\delta-2)k_{d,\delta}(\alpha')]}. 
\label{pia}
\eea 
Therefore,  as long as $\delta<d-2$,  the generalized degree increases dynamically due to an effective "linear preferential attachment" \cite{BA},  according to which the generalized degree of a 
$\delta$-face increases at each time by one,  with a probability increasing linearly with  the current value of its generalized degree.  Since the preferential attachment is a well-known mechanism for 
generating scale-free distributions,  it follows,    by  putting $\delta=0$,   that we expect that as long as  $d\geq3$ the CQNMs  are scale-free.  Instead, in the case $d=2$, by putting $\delta=0$ it is immediate to see that  the probability $\pi_{2, 0}(\alpha)$ is independent of the generalized degree $k_{2,0}(\alpha)$ of the $0-$face (node) $\alpha$, and therefore there is no "effective preferential attachment". We expect therefore \cite{BA_review} that the CQNM in $d=2$ has an exponential degree distribution, i.e. in $d=2$ we expect to observe homogeneous  CQNM with bounded  fluctuations in the degree distribution. 
These arguments can be made rigorous by solving the master equation \cite{Doro_book},  and deriving the exact asymptotic generalized degree distributions for every $\delta<d$ (see  Supplementary Information for details). 
For $\delta=d-1$ we find a bimodal distribution
\bea
P_{d, d-1}(k)=\left\{\begin{array}{lll}\frac{d-1}{d} &\mbox{for} & k=1,  \\
\frac{1}{d} &\mbox{for} & k=2. \end{array}\right. 
\label{d-10}
\eea
For $\delta=d-2$  instead, we find an exponential distribution,  i.e. 
\bea
\begin{array}{ccccc}
P_{d, d-2}(k)&=&\left(\frac{2}{d+1}\right)^{k}\frac{d-1}{2}, & \mbox{for} & k\geq 1. 
\end{array}
\label{d-20}
\eea
Therefore in $d=2$, the CQNMs have an exponential degree distribution that can be derived from Eq. $(\ref{d-20})$ and Eq. $(\ref{Pk})$.
Finally for $0\leq\delta<d-2$ we have the distribution 
\bea
\begin{array}{ccccc}
P_{d, \delta}(k)&=&\frac{d-1}{d-\delta-2}\frac{\Gamma[1+(d+1)/(d-\delta-2)]}{\Gamma[1+2/(d-\delta-2)]}\frac{\Gamma[k+2/(d-\delta-2)]}{\Gamma[k+1+(d+1)/(d-\delta-2)]}, & \mbox{for} &  k\geq 1. 
\end{array}
\label{Pksf}
\eea
It follows that for  $0\leq\delta<d-2$ and $k\gg 1+(d+1)/(d-\delta-2)$ the generalized degree distribution follows a power-law with exponent $\gamma_{d, \delta}$,  i.e. 
\bea 
P_{d, \delta}(k)\simeq Ck^{-\gamma_{d, \delta}}& \mbox{for} &d-\delta>2, 
\eea
and
\bea
\gamma_{d, \delta}=1+\frac{d-1}{d-\delta-2}. 
\label{gammadd}
\eea
The distribution $P_{d, \delta}(k)$ given by Eq.~$(\ref{Pksf})$ is scale-free if an only if $\gamma_{d, \delta}\in(2, 3]$. 
Using Eq.~$(\ref{gammadd})$ we observe that for  $d \geq 3$ and $\delta=0$ we observe that the distribution  of generalized degrees $P_{d, \delta}(k)$ is always scale-free. 
Therefore the degree distribution $P_d(K)$ given by Eq.~$(\ref{Pk})$,  for large values of the degree $K$ and for $d\geq 3$ is  scale-free and  goes like
\bea
P_d(K)\simeq C K^{-\gamma_d}
\eea
with 
\bea
\gamma_d=\frac{2d-3}{d-2}. 
\eea
Therefore, for $d=3$ the CQNMs have $\gamma_3=3$ and for $d\to \infty$ they have power-law exponent $\gamma_d\to 2$.

\begin{figure}
\begin{center}
{\includegraphics[width=1\columnwidth]{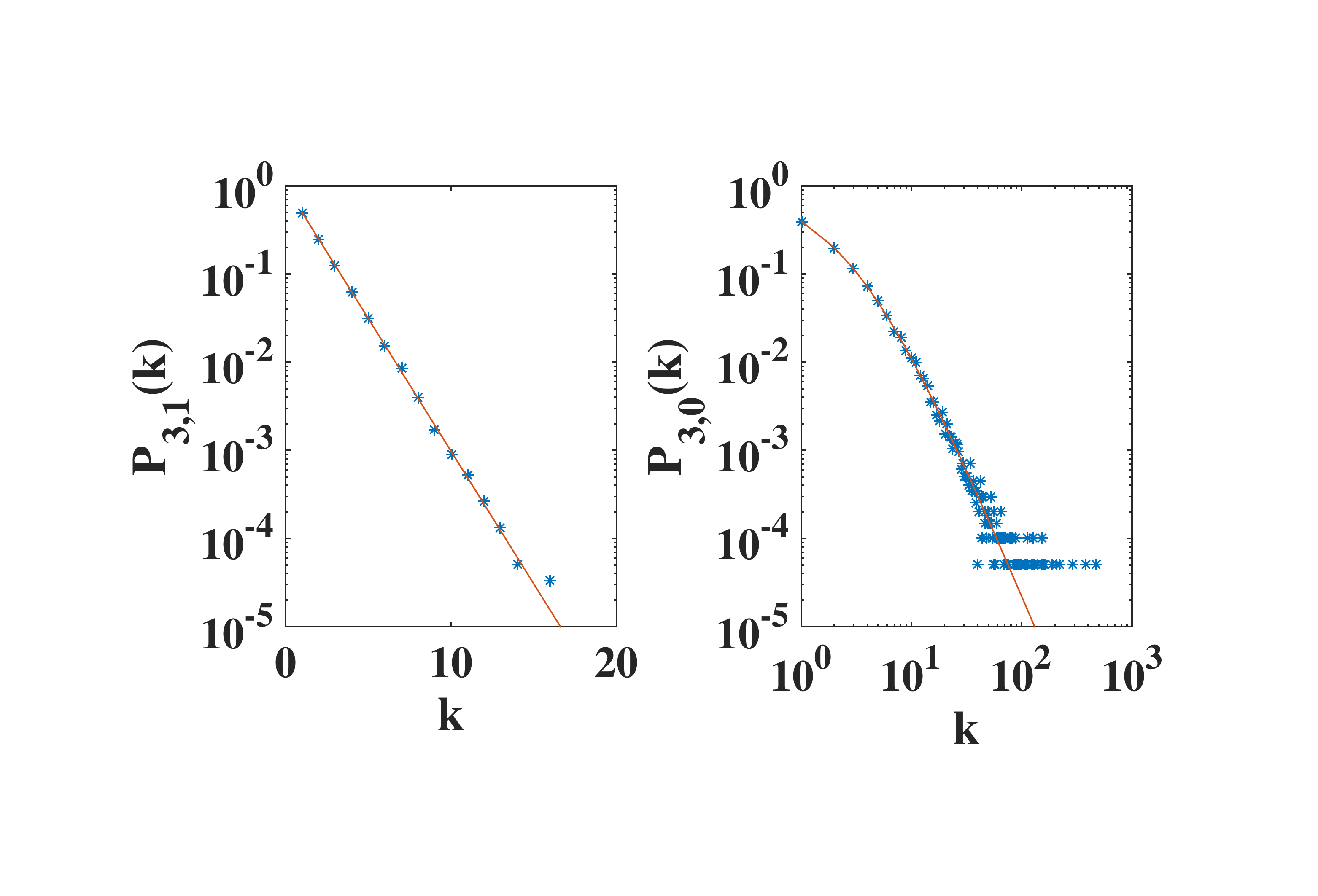}}
\end{center}
\caption{The distribution of the (non-trivial) generalized degrees $P_{3, 1}(k)$ and $P_{3, 0}(k)$ in dimension $d=3$ are shown.  The star symbols indicate the simulation results while the solid red line
indicates the theoretical expectations given respectively by Eqs.  $(\ref{d-20})$ and $(\ref{Pksf})$.  In particular we observe that $P_{3, 1}(k)$ is exponential while $P_{3, 0}(k)$ is scale-free 
implying that the CQNM in $d=3$ is scale-free. The simulation results are shown for a single realization of the CQNM with a total number of nodes $N=2\times 10^4$. }
\label{fig3}
\end{figure}

These theoretical expectations perfectly fit the simulation results of the model  as can been seen in Figure $\ref{fig3}$ where the distribution of generalized degrees $P_{3, 1}(k)$ and $P_{3, 0}(k)$ observed in the simulations for $\beta=0$ are compared with the theoretical expectations. 

In the case $\beta>0$ the distributions of the generalized degrees  depend on the density $\rho_{d, \delta}(\epsilon)$ of $\delta$-dimensional simplices with energy $\epsilon$ in a CQNM and are parametrized by self-consistent parameters called the chemical potentials,  indicated as $\mu_{d, \delta}$ and defined in the  Supplementary Information. 

Here we suppose that these chemical potentials $\mu_{d, \delta}$ exist and that the density $\rho_{d, \delta}(\epsilon)$ is given,  and we find the  self-consistent equations that they need to satisfy  at the end of the derivation. 
Using the master equation approach \cite{Doro_book} we obtain that for   $\delta=d-1$  the generalized degree follows the distribution
\bea
P_{d, d-1}(k)=\left\{\begin{array}{lll}\rho_{d, d-1}(\epsilon)\left[1-\frac{1}{e^{\beta\left(\epsilon-\mu_{d, d-1}\right)}+1}\right],  &\mbox{for} & k=1 \\
\rho_{d, d-1}(\epsilon)\frac{1}{e^{\beta\left(\epsilon-\mu_{d, d-1}\right)}+1} &\mbox{for} & k=2, \end{array}\right. 
\label{d-1e}
\eea
while for $\delta=d-2$ it follows
\bea
\begin{array}{ccccc}
P_{d, d-2}(k)&=&\sum_{\epsilon}\rho_{d, d-2}(\epsilon)\frac{e^{\beta\left(\epsilon-\mu_{d, d-2}\right)}}{\left(e^{\beta\left(\epsilon-\mu_{d, d-2}\right)}+1\right)^{k}}, & \mbox{for} & k\geq 1. 
\end{array}
\label{d-2e}
\eea
Finally for $0\leq\delta< d-2$ the generalized degree is given by 
\bea
\begin{array}{ccccc}
\hspace*{-7mm}P_{d, \delta}(k)&=&\sum_{\epsilon}\rho_{d, d-2}(\epsilon) e^{\beta(\epsilon-\mu_{d, \delta})}\frac{\Gamma\left[1+2/(d-\delta-2)+\exp\left[{\beta\left(\epsilon-\mu_{d, \delta}\right)}\right]\right]\Gamma\left[k+2/(d-\delta-2)\right]}{\Gamma\left[1+2/(d-\delta-2)\right]\Gamma\left[k+1+2/(d-\delta-2)+\exp\left[{\beta\left(\epsilon-\mu_{d, \delta}\right)}\right]\right]}, & \mbox{for} &  k\geq 1. 
\end{array}
\label{deltae}
\eea
It follows that   also for $\beta>0 $ the CQNMs in $d>2$ are scale-free. 
Interestingly,  we observe that the average of the generalized degrees of simplices with energy $\epsilon$ follows the Fermi-Dirac distribution for $\delta=d-1$,  the Boltzmann distribution for $\delta=d-2$ and the Bose-Einstein distribution for $\delta<d-2$. 
In fact we have, 
\bea
\begin{array}{lllr}
{\Avg{[k_{d, d-1}-1]|\epsilon}}&=&n_F(\epsilon, \mu_{d, d-1}), &\  \\
{\Avg{[k_{d, d-2}-1]|\epsilon}}&=&n_Z(\epsilon, \mu_{d, d-2}), &  \  \\
{\Avg{[k_{d, \delta}-1]|\epsilon}}&=&A n_B(\epsilon, \mu_{d, \delta}), & \mbox{for} \ \  \delta<d-2. 
\end{array}
\label{PSB}
\eea
where
$A={(d-\delta)}/{(d-\delta-2)}$,  $n_Z(\epsilon, \mu_{d, d-2})$ is proportional to the Boltzmann distribution and $n_F(\epsilon, \mu_{d, d-1})$, $n_B(\epsilon, \mu_{d, \delta})$ indicate respectively the Fermi-Dirac  and Bose-Einstein occupation numbers \cite{statmech}.  In particular we have
\bea
n_Z(\epsilon, \mu_{d, d-2})&=&e^{-\beta (\epsilon-\mu_{d, d-2})},\nonumber \\
n_F(\epsilon, \mu_{d, d-1})&=&\frac{1}{e^{\beta (\epsilon-\mu_{d, d-1})}+1}, \nonumber \\
n_B(\epsilon,  \mu_{d, \delta})&=&\frac{1}{e^{\beta (\epsilon-\mu_{d, \delta})}-1}\ .
\eea
These results suggest that the dimension $d=3$ of CQNM is the minimal one necessary for observing at the same time scale-free CQNMs and the simultaneous emergence of the Fermi-Dirac,  Boltzmann and Bose-Einstein distributions.  In particular in $d=3$ the average generalized degree of triangles of energy $\epsilon$ follows the Fermi-Dirac distribution,  the average of the generalized degree of links of energy $\epsilon$ follows the Boltzmann distribution,  while the generalized degree of nodes of energy $\epsilon$ follows the Bose-Einstein distribution. 
\begin{figure*}
\begin{center}
{\includegraphics[width=1.1\columnwidth]{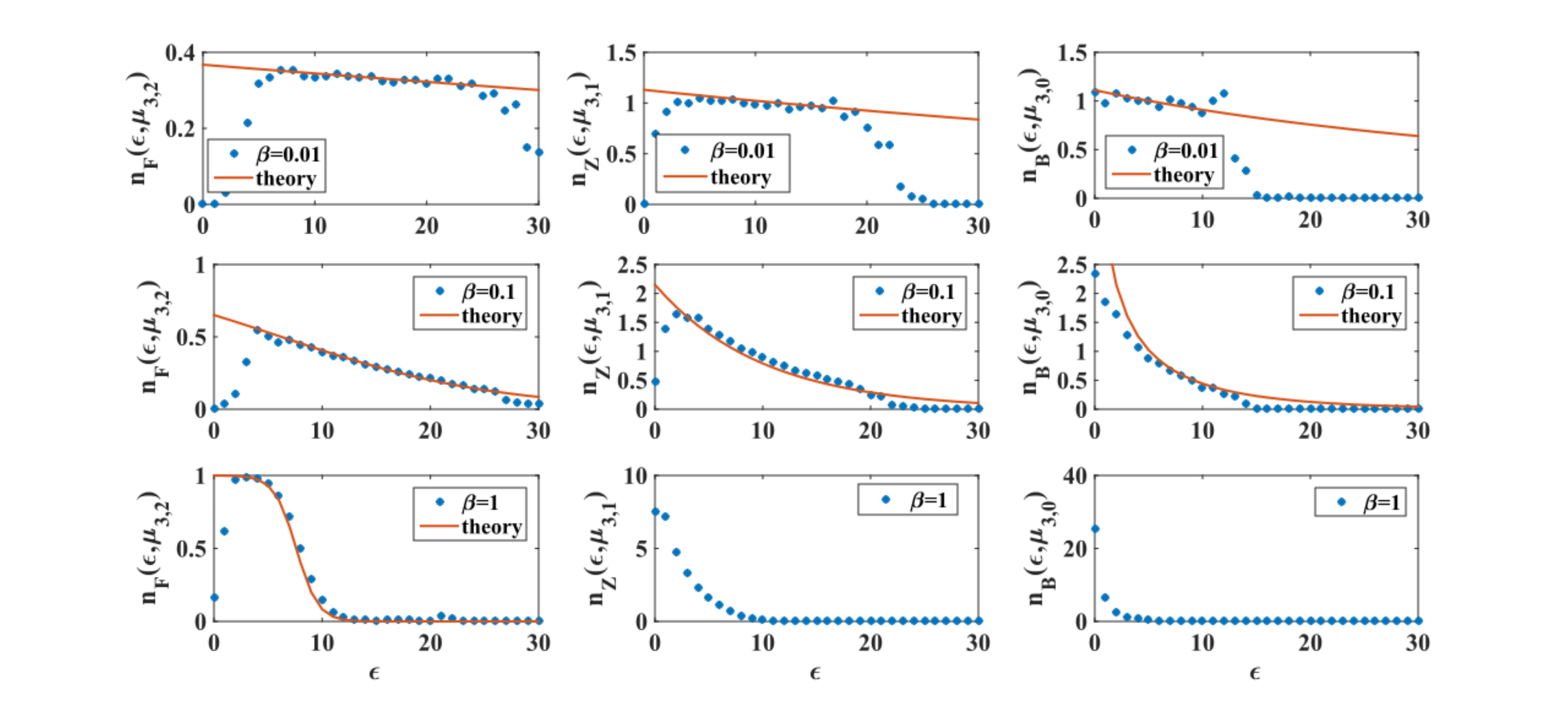}}
\end{center}
\caption{The average of the generalized degrees of $\delta$-faces of energy $\epsilon$, in a CQNM of dimension $d=3$,  follow the Fermi-Dirac $n_F(\epsilon,\mu_{3,2})$, the Boltzmann 
$n_B(\epsilon,\mu_{3,1})$ or the Bose-Einstein distribution $n_B(\epsilon,\mu_{3,0})$ according to Eqs. $(\ref{F})-(\ref{B})$ as long as the chemical potential $\mu_{d,\delta}$ is well-defined,
i.e. for sufficiently low value of the inverse temperature $\beta$. Here we compare simulation results over CQNM of $N=10^3$ nodes in $d=3$ and theoretical results for $\beta=0.01,0.1,1$. The CQNMs in the figure have a Poisson energy distribution $g(\epsilon)$ with average  $z=5$. The simulation results are averaged over $50$ CQNM realizations.}
\label{fig4}
\end{figure*}
Finally the  chemical potentials $\mu_{d, \delta}$,  if they exist,   can be found self-consistently by imposing the condition
\bea
\lim_{t\to \infty}\frac{\sum_{\alpha}k_{d, \delta}}{N_{d, \delta}}=\frac{d+1}{\delta+1}, 
\eea
dictated by the geometry of the CQNM, 
which implies the following self-consistent relations for the chemical potentials $\mu_{d,\delta}$
\bea
&&\int d\epsilon \rho_{d, d-1}(\epsilon) n_F(\epsilon,  \mu_{d, d-1})=\frac{1}{d},  \label{F} \\
&&\int d\epsilon \rho_{d, d-1}(\epsilon) n_Z(\epsilon,  \mu_{d, d-2})
=\frac{2}{d-1},   \ \label{Z}\\
&&\int d\epsilon \rho_{d, \delta}(\epsilon) n_B(\epsilon, \mu_{d, \delta})=\frac{d-\delta-2}{\delta+1},  \  \mbox{for} \ \  \delta<d-2 \label{B}. 
\eea
In Figure $\ref{fig4}$ we compare the simulation results with the theoretical predictions given by Eqs.~$(\ref{PSB})$ finding very good agreement for sufficiently low values of the inverse temperature $\beta$.
The disagreement occurring at large value of the inverse temperature $\beta$ is due to the fact that the  self-consistent Eqs. $(\ref{F})-(\ref{B})$ do not always give a solution for the chemical 
potentials $\mu_{d, \delta}$.  In particular the CQNM with $d\geq 3$ can undergo a Bose-Einstein condensation when Eq.  $(\ref{B})$ cannot be satisfied.  When the transition occurs for the generalized degree with  $\delta=0$,  the maximal degree in the network increases linearly in time similarly to the scenario described in \cite{Bose}.

In summary, we have shown that Complex Quantum Manifolds in dimension $d>2$ are scale-free,  i. e.  they are characterized by large fluctuations of the degrees of the nodes. 
Moreover the $\delta$-faces with  $\delta<d$ follow the Fermi-Dirac,  Boltzmann or Bose-Einstein distributions depending on the dimensions $d$ and $\delta$. In particular for $d=3$,  we find that triangular faces follow the Fermi-Dirac distribution,  links follow the Boltzmann distribution and nodes follow the Bose-Einstein distribution. 
Interestingly, we observe that the dimension $d=3$ is not only the minimal dimension for having a scale-free CQNM,  but it is also the minimal dimension for observing the simultaneous emergence of 
the Fermi-Dirac,  Boltzmann or Bose-Einstein distributions in CQNMs.


\clearpage
\renewcommand\theequation{{S-\arabic{equation}}}
\renewcommand\thetable{{S-\Roman{table}}}
\renewcommand\thefigure{{S-\arabic{figure}}}
\setcounter{equation}{0}
\setcounter{figure}{0}
\setcounter{section}{0}

\onecolumngrid
\appendix

\section*{\Large SUPPLEMENTARY INFORMATION }
\section*{INTRODUCTION}
In this supplementary information we give the details of the derivation discussed in the main text.
In Sec. II we define  Complex Quantum Network Manifolds (CQNMs); in Sec. III we discuss the relation between the CQNM and the evolution of quantum network states; finally in Sec. IV we define the generalized degrees, and we derive the generalized degree distribution in the case $\beta=0$ and $\beta>0$.

\section*{COMPLEX QUANTUM NETWORK MANIFOLDS}
Here we present the non-equilibrium dynamics of  Complex Quantum Network Manifolds (CQNMs). This dynamics is   inspired by biological evolution and self-organized models and generates discrete manifolds formed by simplicial complexes of dimension $d$.
In particular  CQNMs are formed by gluing $d$-simplices along $(d-1)$-faces, in order that each $(d-1)$-face belongs at most to two $d$-dimensional simplices. 

Let us   indicate with ${\cal S}_{d,\delta}$ the set of all $\delta$-faces with $\delta<d$ belonging to the $d$-dimensional CQNMs.
A $(d-1)$-face $\alpha\in {\cal S}_{d,d-1}$ is "saturated" if it belongs to two simplices of dimension $d$, whereas  it is "unsaturated" if it belongs only to a single $d$-dimensional simplex.
We will assign a variable  $\xi_{\alpha}=0,1$  to each   face $\alpha\in {\cal S}_{d,d-1}$, indicating either that the face is unsaturated $(\xi_{\alpha}=1$) or that the face is saturated ($\xi_{\alpha}=0$).

Moreover, to each node $i$ we assign an {\it energy }  $\epsilon_i$ drawn  from a distribution $g(\epsilon)$ and quenched  during the evolution of the network.
To every  $\delta$-face  $\alpha\in {\cal S}_{d,\delta}$ we associate an {\it energy} $\epsilon_{\alpha}$ given by the sum of the energy of the nodes that belong to $\alpha$, 
\bea
\epsilon_{\alpha}=\sum_{i \subset \alpha}\epsilon_i\ .
\eea
At time $t=1$ the CQNM is formed by a single $d$-dimensional simplex.
At  each time $t>1$ we add a simplex of dimension $d$ to an unsaturated $(d-1)$-face $\alpha\in{\cal S}_{d,d-1}$ chosen with probability $\Pi_{\alpha}$ given by  
\bea
\Pi_{\alpha}&=&\frac{1}{Z}e^{-\beta \epsilon_{\alpha}}\xi_{\alpha},
\label{SP1}
\eea
where $\beta$ is a parameter of the model called {\it inverse temperature} and $Z$ is a normalization sum given by 
\bea
Z=\sum_{\alpha\in {\cal S}_{d,d-1}}e^{-\beta \epsilon_{\alpha}}\xi_{\alpha}.
\eea
Having chosen the $(d-1)$-face $\alpha$, we glue to it a new $d$-dimensional complex containing  all the nodes of the face $\alpha$  plus the new node $i$. It follows that the new node $i$ is 
linked to each node $j$ belonging to $\alpha$.\\
Since at time $t=1$ the number of nodes in the CQNM is $N(1)=d+1$, and at each time we add a new additional node, the total number of nodes is $N(t)=t+d$.
The CQNM evolution up to time $t$ is fully determined  by the sequences $\{\epsilon_{i}\}_{i\leq t+d},\{\alpha_{t'}\}_{t'\leq t}$, where $\epsilon_{t'+d}$ indicates the energy of the 
node added to the CQNM at time $t'>1$, $\epsilon_i$ with $i\leq d+1$ indicates the energy of an initial node $i$ of the CQNM, and   $\alpha_{t'}$ indicates  the $(d-1)$-face to which the new $d$-dimensional complex is added at time $t'$.
A similar dynamics for simplicial complexes of dimension $d=2$ has  been proposed in \cite{Emergent}.

\section*{QUANTUM NETWORK STATES}
\subsection*{The network Hilbert space}
Following an  approach similar to  the one used in "Quantum Graphity" and related models \cite{graphity_rg,graphity,graphity_Hamma,Q}, in this section we associate   an Hilbert space ${\cal H}_{tot}$ to a simplicial complex of $N$ nodes formed by gluing together $d$-dimensional simplices along $(d-1)$-faces.
The Hilbert space ${\cal H}_{tot}$ is given by 
\bea
{\cal H}_{tot}=\bigotimes^N {\cal H}_{node} \bigotimes^{P}{\cal H}_{d,d-1} \bigotimes^{P} \tilde{{\cal H}}_{d,d-1},
\eea
with $P=\left(\begin{array}{c}N \nonumber \\
d\end{array}\right)$ indicating  the maximum number of $(d-1)$-faces in a network of   $N$ nodes.
Here an Hilbert space ${\cal H}_{node}$ is associated to each possible node $i$ of the simplicial complex, and two Hilbert spaces ${\cal H}_{d,d-1}$ and $\tilde{{\cal H}}_{d,d-1}$ are associated to each possible  $(d-1)-$face of a network of $N$ nodes. The  Hilbert space ${\cal H}_{node}$  is  the one of   a fermionic oscillator of energy $\epsilon$, with basis 
$\{\ket{o_i,\epsilon}\}$, with $o_i=0,1$.  We indicate with  $b_i^{\dag}(\epsilon),b_i(\epsilon)$ respectively the fermionic creation and annihilation operators acting on this space.
The Hilbert space  ${\cal H}_{d,d-1}$   associated to a $(d-1)$-face  $\alpha$ is the Hilbert space of a  fermionic oscillator with basis
 $\{\ket{a_{\alpha}}\}$, with $a_{\alpha}=0,1$.  We indicate with  $c_{\alpha}^{\dag},c_{\alpha}$ respectively the fermionic creation and annihilation operators acting on this space.
 Finally the  Hilbert space  $\tilde{\cal H}_{d,d-1}$   associated to a $(d-1)$-face $\alpha$ is the Hilbert space of a  fermionic oscillator with basis
 $\{\ket{n_{\alpha}}\}$, with $n_{\alpha}=0,1$. We indicate with  $h_{\alpha}^{\dag},h_{\alpha}$ respectively the fermionic creation and annihilation operators acting on this space. \\
 A quantum network state can therefore be decomposed as
\bea
\ket{\psi(t)}=\sum_{\{o_i,\epsilon_i,a_{\alpha},n_{\alpha}\}} C(\{o_i,\epsilon_i,a_{\alpha},n_{\alpha}\})\prod_i \ket{o_i,\epsilon_i}\prod_{\alpha\in{\cal Q}_{d,d-1}(N)}\ket{a_{\alpha}}\ket{n_{\alpha}},
\eea
where with ${\cal Q}_{d,d-1}(N)$ we indicate all the possible $(d-1)$-faces of a  network of $N$ nodes.\\

 The node states  $\ket{o_i,\epsilon}$ are  mapped respectively to the presence  ($\ket{o_i=1,\epsilon}$) or the absence ($\ket{o_i=0,\epsilon}$) of a node $i$ of energy $\epsilon_i=\epsilon$ in the simplicial complex.
 The quantum state $\ket{a_{\alpha}=1}$ is mapped to the presence of the $(d-1)-$face  $\alpha\in {\cal S}_{d,d-1}$ in the network while the quantum state $\ket{a_{\alpha}=0}$ is mapped 
 to the absence of such a face. 
 Moreover, when $a_{\alpha}=1$, 
 the quantum number  $n_{\alpha}=1$ is mapped to a  saturated $(d-1)-$face  $\alpha$,  i.e.  $\alpha$ is incident to two $d$-dimensional simplices, while the quantum number $n_{\alpha}=0$ is mapped either to an unsaturated 
 $(d-1)-$face  $\alpha$ (if also $a_{\alpha}=1$) or to the absence of such a face (if $a_{\alpha}=0$).

\subsection*{Markovian evolution of the quantum network states}

As already proposed  in the literature \cite{Q, graphity_rg},
here we assume that the quantum network state follows a Markovian evolution. In particular we assume that at time $t=1$ the state is given by 
\bea
\ket{\psi(1)}=\frac{1}{\sqrt{{\cal Z}(1)}}\sum_{\{\epsilon_i\}_{i=1,..d+1}}\prod_{i=1}^{d+1}\sqrt{g(\epsilon_i)}b_i^{\dag}(\epsilon_i)\prod_{\alpha\in{\cal Q}_{d,d-1}(d+1)}c^{\dag}_{\alpha}\ket{0},
\eea 
where ${\cal Z}(1)$ is fixed by the normalization condition $\Avg{\psi(1)|\psi(1)}=1$.
The  quantum network state  is updated at each time $t>1$ according to the unitary transformation 
\bea
\ket{\psi(t)}=U_t\ket{\psi(t-1)}
\label{Smarkov}
\eea
with the unitary operator $U_t$ given by 
\bea
U_t=\sqrt{\frac{{\cal Z}(t-1)}{{\cal Z}(t)}}\sum_{\epsilon_{t+d}}\sqrt{g(\epsilon_{t+d})}b^{\dag}_{t+d}(\epsilon_{t+d})\sum_{\alpha\in {\cal Q}_{d,d-1}(t+d-1)}e^{-\beta \epsilon_{\alpha}/2}\left[\prod_{\alpha' \in {\cal F}(t+d,\alpha)}c^{\dag}_{\alpha'}\right]h^{\dag}_{\alpha}c^{\dag}_{\alpha}c_{\alpha}
\label{SUdef}
\eea
where  ${\cal F}(i,\alpha)$ indicates the set of all the $(d-1)$-faces $\alpha'$ formed by the node $i$ and a subset of the nodes in $\alpha\in {\cal Q}_{d,d-1}(N)$ and  ${\cal Z}(t)$ is fixed by the normalization condition
\bea
\langle \psi(t)|\psi(t)\rangle=1.
\label{Snormdef}
\eea
\subsection*{Path integral characterizing the quantum network state evolution}
The quantity ${\cal Z}(t)$  is a path integral over CQNM evolutions determined by the sequences $\{\epsilon_{i}\}_{i\leq t+d},\{\alpha_{t'}\}_{t'\leq t}$.
In fact, using the normalization condition  in Eq. $(\ref{Snormdef})$ and the evolution of the quantum network state given by Eqs. $(\ref{Smarkov})$, $(\ref{SUdef})$ we get 
\bea
{\cal Z}=\sum_{\{\epsilon_{i}\}_{i \leq t+d}}\sum_{\{\alpha_{t'}\}_{t'\leq t}}W(\{\epsilon_{i}\}_{i\leq t+d },\{\alpha_{t'}\}_{t'\leq t})
\eea
where $W(\{\epsilon_{i}\}_{i\leq t+d },\{\alpha_{t'}\}_{t'\leq t})$ is given by  
\bea\label{SWdef}
W(\{\epsilon_{i}\}_{i\leq t+d },\{\alpha_{t'}\}_{t'\leq t})
&=&\prod_{i=1}^{t+d}g(\epsilon_i)\prod_{t'\leq t}a_{\alpha_{t'}}(t')(1-n_{\alpha({t'})}(t'))e^{-\beta\sum_{\alpha\in {\cal Q}_{d,d-1}(t)}\epsilon_{\alpha} n_{\alpha}(t)},
\label{SWdef2}
\eea
where the terms  $a_{\alpha}(t)$ and $n_{\alpha}(t)$ that appear in Eq. $(\ref{SWdef})$ can be expressed in terms of the history $\{\alpha_{t'}\}_{t'\leq t}$ as
\bea\label{SFDaandn}
a_{\alpha}(t)&=&\sum_{t'= 2}^{t-1}\left(\sum_{\tilde{\alpha} \in {\cal F}(\alpha_{t'},t'+d)}\delta[\alpha,\tilde{\alpha}]\right)+\sum_{\tilde{\alpha} \in {\cal Q}_{d,d-1}(d+1)}\delta[\alpha,\tilde{\alpha}],\nonumber \\
 n_{\alpha}(t)&=&\sum_{t'=2}^{t-1}\delta[\alpha_{t'},\alpha].
 \eea

We note that ${\cal Z}(t)$ can also be interpreted  as the partition function of the statistical mechanics problem over possible evolutions of CQNM.
In fact, the  CQNM evolution is determined by the sequences $\{\epsilon_{i}\}_{i\leq t+d },\{\alpha_{t'}\}_{t' \leq t}$
 where $\epsilon_{t'+d}$ indicates the energy of the node added at time $t'>1$ and $\epsilon_i$ for $i \leq d+1$ indicates the energy of a node $i$  in the CQNM at time $t=1$, and $\alpha_{t'}$ indicates the $(d-1)$-face to which we attach the new $d$-dimensional simplex at time $t'$.
The probability $P(\{\epsilon_{i}\}_{i\leq t+d },\{\alpha_{t'}\}_{t' \leq t})$ of a given evolution is given by  
\bea
P(\{\epsilon_{i}\}_{i\leq t+d },\{\alpha_{t'}\}_{t' \leq t})=\frac{W(\{\epsilon_{i}\}_{i \leq t+d },\{\alpha_{t'}\}_{t' \leq t})}{{\cal Z}(t)}
\eea
where $W(\{\epsilon_{i}\}_{i \leq t+d },\{\alpha_{t'}\}_{t'\leq t})$ is given by Eq. $(\ref{SWdef2})$ and ${\cal Z}(t)$ is fixed  by the condition Eq. $(\ref{Snormdef})$.
This implies that the set of all  classical evolutions of the CQNM fully determine the properties of the quantum network state evolving through the Markovian dynamics given by Eq. $(\ref{Smarkov})$.

\section*{GENERALIZED DEGREES}
\subsection*{Definition of generalized degrees}
A set of  important structural properties of the CQNM are the generalized degrees $k_{d,\delta}(\alpha)$ of the $\delta$-faces  in a $d$-dimensional CQNM.
Given a CQNM of dimension $d$, the generalized degree $k_{d,\delta}(\alpha) $ of a given $\delta$-face $\alpha$, (i.e. $\alpha\in {\cal S}_{d,\delta}$) is defined as the number of $d$-dimensional simplices  incident to it.
If we consider the adjacency tensor ${\bf a}$ of elements $a_{\alpha}=1$ if the $d$-dimensional complex is part of the CQNM and otherwise zero, $a_{\alpha}=0$, the generalized degree  of a $\delta$-face $\alpha'$ is given by 
\bea
k_{d,\delta}(\alpha')=\sum_{\alpha | \alpha' \subset \alpha}a_{\alpha}.
\eea
For example, in a CQNM of dimension $d=2$, the generalized degree $k_{2,1}(\alpha)$ is the number of triangles incident to a link $\alpha$ while the generalized degree $k_{2,0}(\alpha)$ indicates the number of triangles incident to a node $\alpha$.
Similarly in a CQNM of dimension $d=3$, the generalized degrees $k_{3,2}$, $k_{3,1}$ and $k_{3,0}$ indicate  the number of tetrahedra incident respectively to a triangular face, a link or a node.
\subsection*{Distribution of Generalized Degrees for $\beta=0$}

Let us define  the probability $\pi_{d,\delta}(\alpha)$ that a new $d$-dimensional simplex is attached to a $\delta$-face $\alpha$.
Since each $d$-dimensional simplex is attached to a random unsaturated $(d-1)-$face, and the number of such faces is $(d-1)t$, we have that for $\delta=d-1$ 
\bea
\pi_{d,d-1}(\alpha)=\left\{\begin{array}{ccc}\frac{1}{(d-1)t}& \mbox{for} &k=1,\nonumber \\
0&\mbox{for}& k=2. \end{array}\right.
\label{Spia0}
\eea 
Let us now observe that each $\delta$-face, with  $\delta<d-1$, which has generalized  degree $k_{d,\delta}(\alpha)=k$, is incident to 
\bea
2+(d-\delta-2)k,
\label{S2d}
\eea
unsaturated $(d-1)$-faces. \\
In fact, is is easy to check that a $\delta-$face  with generalized degree  $k_{d,\delta}=k=1$   is incident to $d-\delta$ unsaturated $(d-1)$-faces. 
Moreover, at each time we add to a $\delta$-face a new $d$-dimensional simplex, a number $d-\delta-1$ of unsaturated $(d-1)-$faces are added to the $\delta$-face  while a previously unsaturated $(d-1)$-face incident to it  becomes saturated.
Therefore the number of $(d-1)$-unsaturated faces incident to a $\delta$-face of generalized degree $k_{d,\delta}=k$ follows 
Eq. $(\ref{S2d})$.
We have therefore that  the probability 
$\pi_{d,\delta}(\alpha)$ to attach a new $d$-dimensional simplex to a $\delta$-face $\alpha$ with $\delta<d-1$ and generalized degree $k_{d,\delta}(\alpha)$ is given by 
\bea
\pi_{d,\delta}(\alpha)=\frac{2+(d-\delta-2)k_{d,\delta}(\alpha)}{\sum_{\alpha'\in {\cal S}_{d,\delta}}[2+(d-\delta-2)k_{d,\delta}(\alpha')]},
\label{Spia2}
\eea 
where for large times $t\gg1, $ 
\bea 
\sum_{\alpha'\in {\cal S}_{d,\delta}}[2+(d-\delta-2)k_{d,\delta}(\alpha')]=2\left(\begin{array}{c}d \\ \delta \end{array}\right)t+(d-\delta-2)\left(\begin{array}{c}d+1 \\ \delta+1\end{array}\right)t=(d-1)\left(\begin{array}{c}d \\ \delta+1\end{array}\right)t.\nonumber
\eea
From Eq. $(\ref{Spia2})$ if follows that, as long as $\delta<d-2$, the generalized degree follows a "preferential attachment" mechanism  \cite{BA,BA_review, Newman_book,Boccaletti2006,Caldarelli_book,Doro_book}. 

Moreover, the average number $n_{d,\delta}(k)$ of $\delta$-faces of generalized degree $k_{d,\delta}=k$ that increases their generalized degree by one  at a generic time $t>1$, is given by 
\bea
n_{d,\delta}(k)=\left.m_{d,\delta+1}\pi_{d,\delta}(\alpha)\right|_{k_{d,\delta}(\alpha)=k}
\eea
where $m_{d,\delta+1}=\left(\begin{array}{c}d \\ \delta+1\end{array}\right)$ indicates the total number of $\delta-$faces incident to the $d$-dimensional simplex added at time $t$.
This implies that for  $\delta=d-1$, and large times $t\gg1$, $n_{d,\delta}(k)$ is given by 
\bea
n_{d, d-1}(k)=\frac{1}{(d-1)t}\delta_{k, 1}
\label{Spia00}
\eea 
where $\delta_{x, y}$ indicates the Kronecker delta, while for  $\delta<d-1$ it is given by 
\bea
n_{d, \delta}(k)=\frac{2+(d-\delta-2)k}{(d-1)t}.
\label{Spia10}
\eea 
Using Eqs. $(\ref{Spia00})-(\ref{Spia10})$ and the master equation approach \cite{Doro_book}, it is possible to derive the exact distribution for the generalized degrees.
We indicate with $N_{d,\delta}^t(k)$ the average  number of $\delta$-faces that at time $t$ have generalized degree $k_{d,\delta}=k$ during the temporal evolution of a $d$-dimensional CQNM. The master equation \cite{Doro_book} for $N_{d,\delta}^t(k)$ reads
\bea
N^{t+1}_{d,\delta}(k)-N^{t}_{d,\delta}(k)=n_{d,\delta}(k-1)N_{d,\delta}^t(k-1)(1-\delta_{k,1})-n_{d,\delta}(k)N^t_{d,\delta}(k)+m_{d,\delta}\delta_{k,1}
\eea 
with $k\geq 1$.
Here $\delta_{x,y}$ indicates the Kronecker delta, and $m_{d,\delta}=\left(\begin{array}{c}d\nonumber \\ \delta\end{array}\right)$ is the number of  $\delta$-faces added at each time $t$ to the CQNM.
The master equation is solved by observing that  for large times $t\gg1 $ we have $N_{d,\delta}^t(k)\simeq m_{d,\delta}tP_{d,\delta}(k)$ where $P_{d,\delta}(k)$ is the generalized degree distribution.
For $\delta=d-1$ we obtain the  bimodal distribution
\bea
P_{d,d-1}(k)=\left\{\begin{array}{lll}\frac{d-1}{d}, &\mbox{for} & k=1\nonumber \\
\frac{1}{d} &\mbox{for} & k=2\end{array}\right.\ .
\eea
For $\delta=d-2$ instead, we find an exponential distribution, i.e.
\bea
\begin{array}{ccccc}
P_{d,d-2}(k)&=&\left(\frac{2}{d+1}\right)^{k}\frac{d-1}{2},& \mbox{for} & k\geq 1 \ .
\end{array}
\label{SPkex}
\eea
Finally for $0\leq\delta<d-2$ we have the distribution 
\bea\label{Spksf}
\begin{array}{ccccc}
P_{d,\delta}(k)&=&\frac{d-1}{d-\delta-2}\frac{\Gamma[1+(d+1)/(d-\delta-2)]}{\Gamma[1+2/(d-\delta-2)]}\frac{\Gamma[k+2/(d-\delta-2)]}{\Gamma[k+1+(d+1)/(d-\delta-2)]},& \mbox{for} &  k\geq 1.
\end{array}
\label{SPksf}
\eea
From Eq. $(\ref{Spksf})$ it follows that for  $0\leq\delta<d-2$ and $k\gg1$ the generalized degree distribution follows a power-law with exponent $\gamma_{d,\delta}$, i.e.
\bea 
P_{d,\delta}(k)\simeq Ck^{-\gamma_{d,\delta}}& \mbox{for} &d-\delta>2,
\eea
and
\bea
\gamma_{d,\delta}=1+\frac{d-1}{d-\delta-2}.
\label{Sgammadd}
\eea
Therefore the generalized degree  distribution $P_{d,\delta}(k)$ given by Eq. $(\ref{SPksf})$ is scale-free, i.e. it has diverging second moment $\Avg{\left(k_{d,\delta}\right)^2}$, as long as  $\gamma_{d,\delta}\in(2,3]$.
This implies that the  generalized degree distribution is scale-free for 
\bea
\delta<\frac{d-3}{2}.
\eea

\subsection*{Distribution of Generalized Degrees for $\beta>0$}
In the case $\beta>0$ the distribution of the generalized degrees $P_{d,\delta}(k)$ are convolutions of the conditional probabilities $P_{d,\delta}(k|\epsilon)$ that $\delta$-faces with energy $\epsilon$ have given generalized degree $k_{d,\delta}=k$.
Here we derive the distribution of generalized degrees $P_{d,\delta}(k)$ for different values of $\delta$ and $d$ as a function of the inverse temperature $\beta$. 
The procedure for finding these distributions is similar for every value of $\delta$. \\
First we will determine the master equations \cite{Doro_book} for the average number $N_{d,\delta}^t(k|\epsilon)$ of $\delta$-faces of energy $\epsilon$ that have generalized degree $k$ at time $t$. 
Then we will solve these equations, imposing the scaling, valid in general for growing networks, given by $N_{d,\delta}(k|\epsilon)=m_{d,\delta}\rho_{d,\delta}(\epsilon)tP_{d,\delta}(k|\epsilon)$, 
where $\rho_{d,\delta}(\epsilon)$ is the probability that a $\delta$-face has energy $\epsilon$ and $m_{d,\delta}$ are the number of $\delta$-faces added at each time to the CQNM. The master equations will also depend on  self-consistent parameters $\mu_{d,\delta}$ called {\it chemical potentials} that need to satisfy self-consistent equations for the derivation to hold.\\
Let us consider first the case $\delta=d-1$. 
The average number $N_{d,\delta}^t(k|\epsilon)$ of $\delta=(d-1)$-faces of energy $\epsilon$ that at  time $t$ have  generalized degrees $k_{d,\delta}=k$ follows the master equation given by 
\bea
{ N_{d,d-1}^{t+1}(k=2|\epsilon)}-N_{d,d-1}^{t}(k=2|\epsilon)&=&\frac{e^{-\beta \epsilon}}{Z}N_{d,d-1}^t({k=1}|\epsilon),\nonumber \\
&&\nonumber \\
{N_{d,d-1}^{t+1}(k=1|\epsilon)}-N_{d,d-1}^{t}(k=1|\epsilon)&=&-\frac{e^{-\beta \epsilon}}{Z}N_{d,d-1}^t(k=1|\epsilon)+m_{d,d-1}\rho_{d,d-1}(\epsilon)
\label{SNFn}
\eea
where  $\rho_{d,d-1}(\epsilon)$ is the probability that a    $(d-1)$-face  added to the network at a generic time $t\gg1$ has energy $\epsilon$ and $m_{d,d-1}=\left(\begin{array}{c} d\nonumber \\ d-1 \end{array}\right)=d$ is the number of $\delta$-faces added to the network at each time $t$.
In order to solve this master equation we assume that the normalization constant $Z\propto t$ and we put
\bea
e^{-\beta \mu_{d,d-1}}&=&\lim_{t\to \infty}\frac{Z}{t}.
\label{SselfF0}
\eea
This is a self-consistent assumption that must be verified by the solution of Eqs. $(\ref{SNFn})$.
Moreover we observe  that at large times $N_{d,d-1}^t(k|\epsilon)\simeq m_{d,d-1}\rho_{d,d-1}(\epsilon) P_{d,d-1}(k|\epsilon)$. 
Here $P_{d,d-1}(k|\epsilon)$ indicates the asymptotic probability that a $(d-1)$-face $\alpha$ with energy $\epsilon$ has $k_{d,d-1}(\alpha)=k$. With these assumptions, 
we can solve Eqs. $(\ref{SNFn})$ finding
\bea
\hspace*{-10mm}P_{d,d-1}(k=1|\epsilon)&=&\rho_{d,d-1}(\epsilon)\frac{e^{\beta(\epsilon-\mu_{d,d-1})}}{e^{\beta(\epsilon-\mu_{d,d-1})}+1}=\rho_{d,d-1}(\epsilon)[1-n_F(\epsilon, \mu_{d,d-1})]\nonumber \\
\hspace*{-10mm}P_{d,d-1}(k=2|\epsilon)&=&\rho_{d,d-1}(\epsilon)\frac{1}{e^{\beta(\epsilon-\mu_{d,d-1})}+1}=\rho_{d,d-1}(\epsilon)n_F(\epsilon,\mu_{d,d-1}),
\label{SNsF}
\eea
where $n_F(\epsilon,\mu_{d,d-1})$ is the Fermi-Dirac occupation number with chemical potential $\mu_{d,d-1}$, i.e.
\bea
n_F(\epsilon,\mu_{d,d-1})=\frac{1}{e^{\beta(\epsilon-\mu_{d,d-1})}+1}.
\eea
Using Eqs. $(\ref{SNsF})$, and performing the average $\Avg{k_{d,d-1}|\epsilon}$ of the generalized degree $k_{d,d-1}$ over 
$(d-1)$-faces of energy $\epsilon$, one can easily find that 
\bea
\Avg{[k_{d,d-1}-1]|\epsilon}=\frac{1}{e^{\beta (\epsilon-\mu_{d,d-1})}+1}=n_F(\epsilon,\mu_{d,d-1}).
\eea
This result shows that the average  of generalized degrees of $(d-1)$-faces of energy $\epsilon$ is determined by the Fermi-Dirac statistics with chemical potential $\mu_{d,d-1}$.
\\
Let us now consider the case $\delta=d-2$.
In this case we assume   that asymptotically in time we can define the chemical potential ${\mu}_{d,d-2}$ as 
\bea
\hspace*{-10mm}e^{\beta {\mu}_{d,d-2}}=e^{\beta \mu_{d,d-1}}\lim_{t \to \infty} \Avg{\frac{\sum_{\alpha \in{\cal Q}_{d,d-2}(t)}\sum_{\alpha'\in {\cal Q}_{d,d-1}(t)| \alpha\subset \alpha'}e^{-\beta (\epsilon_{\alpha'}-\epsilon_{\alpha})}\xi_{\alpha'}\delta(k_{d,d-2}(\alpha),k)}{\sum_{\alpha \in{\cal Q}_{d,d-2}(t)}\delta(k_{d,d-2}(\alpha),k)}}_{k}.
\label{Sselfd_2}
\eea
In this assumption, the master equations \cite{Doro_book} for the average number $N^t_{d,d-2}(k|\epsilon)$ of $(d-2)$-faces with energy $\epsilon$ and generalized degree $k\geq 1$, read  
\bea
\hspace*{-0mm}{N_{d,d-2}^{t+1}(k|\epsilon)}-N_{d,d-2}^{t}(k | \epsilon)&=&\frac{e^{-\beta (\epsilon-{\mu}_{d,d-2})}}{t}N_{d,d-2}^t(k-1|\epsilon)[1-\delta_{k,1}]-\frac{e^{-\beta (\epsilon-{\mu}_{d,d-2})}}{t}N_{d,d-2}^t(k|\epsilon)\nonumber \\
&&+m_{d,d-2}\rho_{d,d-2}(\epsilon)\delta_{k,1},
\label{SNZk}
\eea
where $m_{d,d-2}=d(d-1)/2$ is the number of $(d-2)-$faces added at each time $t$ to the CQNM, $\rho_{d,d-2}(\epsilon)$ is the probability that such faces have energy $\epsilon$, and $\delta_{x,y}$ indicates the Kronecker delta.
In the large network limit $t\gg1$ we observe that  $N_{d,d-2}^t(k|\epsilon)\simeq tm_{d,d-2}\rho_{d,d-2}(\epsilon)P_{d,d-2}(k|\epsilon)$ where $P_{d,d-2}(k|\epsilon)$ indicates the probability that a $(d-2)-$face of energy $\epsilon$ has generalized degree $k$. Solving Eq. ($\ref{SNZk}$) we get, 
\bea
P_{d,d-2}(k|\epsilon)=\rho_{d,d-2}(\epsilon)\frac{e^{\beta(\epsilon-{\mu}_{d,d-2})}}{\left[e^{\beta (\epsilon-{\mu}_{d,d-2})}+1\right]^{k}},
\label{SPkFc}
\eea
for $k\geq 1$.
Therefore, summing over all the values of the energy of the nodes $\epsilon$ we get the full degree distribution $P(k)$
\bea
P_{d,d-2}(k)=\sum_{\epsilon}\rho_{d,d-2}(\epsilon)\frac{e^{\beta(\epsilon-{\mu}_{d,d-2})}}{\left[e^{\beta (\epsilon-{\mu}_{d,d-2})}+1\right]^{k}},
\label{SPkZ}
\eea
for $k\geq 1$.
Using Eqs. $(\ref{SPkZ})$, and performing the average $\Avg{k_{d,d-2}|\epsilon}$ of the generalized degree $k_{d,d-2}$ over 
$(d-2)$-faces of energy $\epsilon$, one can easily find that 
\bea
\Avg{[k_{d,d-2}-1]|\epsilon}=e^{-\beta (\epsilon-\mu_{d,d-2})}=n_Z(\epsilon,\mu_{d,d-2}),
\eea
where we have indicated with $n_Z(\epsilon,\mu_{d,d-2})$ the Boltzmann distribution 
\bea
n_Z(\epsilon,\mu_{d,d-2})=e^{-\beta (\epsilon-\mu_{d,d-2})}.
\eea
This result shows that the average  of generalized degrees of $(d-2)$-faces of energy $\epsilon$ is determined by the Boltzmann statistics with chemical potential $\mu_{d,d-2}$.
\\
Let us finally consider  the case $\delta<d-2$.
In this case we assume   that asymptotically in time we can define the chemical potential ${\mu}_{d,\delta}$ given by 
\bea
\hspace*{-10mm}
e^{\beta {\mu}_{d,\delta}}=e^{\beta \mu_{d,d-1}}\lim_{t \to \infty} \Avg{\frac{\sum_{\alpha \in{\cal Q}_{d,\delta}(t)}\sum_{\alpha'\in {\cal Q}_{d,d-1}(t)| \alpha\subset \alpha'}e^{-\beta (\epsilon_{\alpha'}-\epsilon_{\alpha})}\xi_{\alpha'}\delta(k_{d,\delta}(\alpha),k)}{\sum_{\alpha \in{\cal Q}_{d,\delta}(t)}\left[k+2/(d-2-\delta)\right]\delta(k_{d,\delta}(\alpha),k)}}_{k}.\label{Sselfdelta}
\eea
 Assuming that the chemical potential $\mu_{d,\delta}$ exists, the master equations \cite{Doro_book} for the average number $N^t_{d,\delta}(k|\epsilon)$ of $\delta$-faces with energy $\epsilon$ and generalized degree $k\geq 1$ read  
\bea
{N_{d,\delta}^{t+1}(k|\epsilon)}-N_{d,\delta}^{t}(k | \epsilon)&=&\frac{e^{-\beta (\epsilon-{\mu}_{d,\delta})}[k-1+2/(d-\delta-2)]}{t}N_{d,\delta}^t(k-1|\epsilon)[1-\delta_{k,1}]\nonumber \\
&&-\frac{e^{-\beta (\epsilon-{\mu}_{d,\delta})}[k+2/(d-\delta-2)]}{t}N_{d,\delta}^t(k|\epsilon)\nonumber \\
&&+m_{d,\delta}\rho_{d,\delta}(\epsilon)\delta_{k,1},
\label{SNBk}
\eea
where $m_{d,\delta}=\left(\begin{array}{c}d\nonumber \\\delta\end{array}\right)$ is the number of $\delta-$faces added at each time $t$ to the CQNM, $\rho_{d,\delta}(\epsilon)$ is the probability that such faces have energy $\epsilon$, and $\delta_{x,y}$ indicates the Kronecker delta.
In the  large network limit $t\gg1$ we observe that $N_{d,\delta}^t(k|\epsilon)\simeq tm_{d,\delta}\rho_{d,\delta}(\epsilon)P_{d,\delta}(k|\epsilon)$,where $P_{k,\delta}(k|\epsilon)$ is the probability that a $\delta-$face of energy $\epsilon$ has generalized degree $k$. Solving Eq. ($\ref{SNBk}$) we get, 
\bea
\begin{array}{ccccc}
\hspace*{-7mm}P_{d,\delta}(k)&=&\sum_{\epsilon}\rho_{d,d-2}(\epsilon) e^{\beta(\epsilon-\mu_{d,\delta})}\frac{\Gamma\left[1+2/(d-\delta-2)+\exp\left[{\beta\left(\epsilon-\mu_{d,\delta}\right)}\right]\right]\Gamma\left[k+2/(d-\delta-2)\right]}{\Gamma\left[1+2/(d-\delta-2)\right]\Gamma\left[k+1+2/(d-\delta-2)+\exp\left[{\beta\left(\epsilon-\mu_{d,\delta}\right)}\right]\right]},& \mbox{for} &  k\geq 1.
\end{array}
\label{SPkB}
\eea
Using Eqs. $(\ref{SPkB})$, and performing the average $\Avg{k_{d,\delta}|\epsilon}$ of the generalized degree $k_{d,\delta}$ over $\delta$-faces of energy $\epsilon$, one can easily find that 
\bea 
{\Avg{[k_{d,\delta}-1]|\epsilon}}&=A\frac{1}{e^{\beta (\epsilon-\mu_{d,\delta})}-1}=An_B(\epsilon,\mu_{d,\delta}),
\eea
where
$A={(d-\delta)}/{(d-\delta-2)}$ and where $n_B(\epsilon,\mu_{d,\delta})$ indicates the Bose-Einstein occupation number with chemical potential $\mu_{d,\delta}$, i.e.
\bea
n_B(\epsilon,\mu_{d,\delta})=\frac{1}{e^{\beta (\epsilon-\mu_{d,\delta})}-1}.
\eea
This result shows that the average  of generalized degrees of $\delta$-faces of energy $\epsilon$ is determined by the Bose-Einstein statistics with chemical potential $\mu_{d,\delta}$.
Finally the  chemical potentials $\mu_{d,\delta}$, if they exist,  can be found self-consistently by imposing the condition
\bea
\lim_{t\to \infty}\frac{\sum_{\alpha}k_{d,\delta}}{N_{d,\delta}}=\frac{\left(\begin{array}{c}d+1\\ \delta+1\end{array}\right)}{\left(\begin{array}{c}d\\ \delta \end{array}\right)}=\frac{d+1}{\delta+1},
\label{Ss}
\eea
dictated by the geometry of the CQNM.
In fact at each time $t$ we add to the network $m_{d,\delta}=\left(\begin{array}{c}d\\ \delta \end{array}\right)$ new $\delta$-faces
and we increases the sum of the generalized degree $k_{d,\delta}$ by the amount  $\left(\begin{array}{c}d+1\\ \delta+1\end{array}\right)$.
Imposing Eq. $(\ref{Ss})$ implies the following normalization constraints for the chemical potentials $\mu_{d,\delta}$,
\bea
\begin{array}{lllr}
\int d\epsilon \rho_{d,d-1}(\epsilon) n_F(\epsilon, \mu_{d,d-1})=\frac{1}{d},&\ \label{SF} \\
\int d\epsilon \rho_{d,d-1}(\epsilon) n_Z(\epsilon, \mu_{d,d-2})=\frac{2}{d-1},&  \ \label{SZ} \\
\int d\epsilon \rho_{d,\delta}(\epsilon) n_B(\epsilon,\mu_{d,\delta})=\frac{d-\delta-2}{\delta+1},& \mbox{for} \ \  \delta<d-2.\label{SB}
\end{array}
\eea
For small values of $\beta$, these equations have  a solution that converges for $\beta\to 0$ to the $\beta=0$ solution discussed in the previous subsection.
As the value of $\beta$ increases it is possible that the chemical potentials $\mu_{d,\delta}$ become ill-defined and do not exist. In this case different phase transitions can occur. 
For the case $d=2$ these transitions have been discussed in detail in \cite{Q}. For $d>2$ we observe that the network might undergo a Bose-Einstein phase transition for values of the inverse 
temperature for which  Eq. $(\ref{SB})$ cannot be solved in order to find the chemical potential $\mu_{d,\delta}$. The detailed discussion of the possible phase transitions in CQNM is beyond the scope of this work and will be the subject of a separate publication.
\subsection*{Mean-field treatment of the case $\beta>0$}
It is interesting to characterize the evolution in time of the generalized degrees using the mean-field approach \cite{Doro_book}.
This approach reveals other aspects of the model that are responsible for the emergence of the statistics determining the distribution of the generalized degrees.
Let us consider separately the mean-field equations determining the evolution of the generalized degrees of $\delta-$faces with $\delta=d-1,d-2$ or with $\delta<d-2$.\\
The $(d-1)$-faces can have generalized degree $k_{d,d-1}$ that can take only two values $k_{d,d-1}=1,2$.
The indicator $\xi_{\alpha}$ of a  $(d-1)-$face $\alpha$ with generalized degree $k_{d,d-1}(\alpha)=k$ is given by   
\bea
{\xi}_{\alpha}=2-k_{d,d-1}(\alpha).
\eea
In fact for $k=2$ the face is saturated and $\xi_{\alpha}=0$ while for $k=1$ the face is unsaturated, therefore $\xi_{\alpha}=1$.
The mean-field approach consists in neglecting fluctuations, and identifying the variable $\xi_{\alpha}$ (evaluated  at time $t$ for  a $\delta-$face $\alpha$ arrived in the CQNM at time $t_{\alpha}$)  with its average $\xi_{\alpha}=\hat{\xi}_{\alpha}(t,t_{\alpha})$ over all the CQNM realizations.
The mean-field equation  for $\hat{\xi}_{\alpha}$ is given by 
\bea
\frac{d\hat{\xi}_{\alpha}}{dt}=-\frac{e^{-\beta \epsilon_{\alpha}}\hat{\xi}_{\alpha}}{Z},
\label{Sminus}
\eea
with initial condition $\hat{\xi}_{\alpha}(t_{\alpha},t_{\alpha})=1$ where $t_{\alpha}$ is the time at which the $(d-1)-$face is added to the CQNM.
The dynamical Eq. $(\ref{Sminus})$ is derived from the dynamical rules of the CQNM evolution.  In fact, at each time  one $(d-1)-$face is chosen with probability $\Pi_{\alpha}$ given by Eq. $(\ref{SP1})$. This face   becomes unsaturated and glued to the new $d$-dimensional simplex. Therefore at each time  $\hat{\xi}_{\alpha}$  indicating the average of $\xi_{\alpha}$ decreases in time by an amount  given by  $\Pi_{\alpha}$.
Assuming that for large time $t$ we have $Z\simeq e^{-\beta \mu_{d,d-1}}t,$ where the chemical potential $\mu_{d,d-1}$ is defined in Eq. $(\ref{SselfF0})$, it follows that the solution of the mean-field Eq. $(\ref{Sminus})$ is given by 
\bea
\hat{\xi}_{\alpha}(t,t_{\alpha})=\left(\frac{t_{\alpha}}{t}\right)^{e^{-\beta(\epsilon_{\alpha}-\mu_{d,d-1})}}.
\eea
The average of $\hat{\xi}_{\alpha}$ over all  $(d-1)-$faces $\alpha$ with energy $\epsilon$, i.e. $\Avg{\hat{\xi}_{\alpha}|\epsilon}$ is given by 
\bea
\Avg{\hat{\xi_{\alpha}}|\epsilon}=\Avg{[2-k_{d,d-1}]|\epsilon}=\frac{1}{t}\int_{1}^t dt_{\alpha}\left(\frac{t_{\alpha}}{t}\right)^{e^{-\beta(\epsilon-\mu_{d,d-1})}}=\frac{e^{\beta(\epsilon-\mu_{d,d-1})}}{1+e^{\beta(\epsilon-\mu_{d,d-1})}}+{\it o}(t^{-1}).
\eea
Therefore we obtain also in the mean field approximation, that the average generalized degree of $(d-1)-$faces with energy $\epsilon$, for $t\gg1$ satisfies, 
\bea
\Avg{[k_{d,d-1}-1]|\epsilon}=\frac{1}{1+e^{\beta(\epsilon_{\alpha}-\mu_{d,d-1})}}=n_{F}(\epsilon-\mu_{d,d-1}),
\eea
where $n_F(\epsilon,\mu_{d,d-1})$ indicates the Fermi-Dirac occupation number with chemical potential $\mu_{d,d-1}$.\\
Let us consider now the generalized degree of $(d-2)$-faces using the mean-field approximation.
Assuming that  the chemical potential $\mu_{d,d-2}$ defined in Eq. $(\ref{Sselfd_2})$ is well defined, it is possible to write down the mean field equation for  the average $\hat{k}_{d,d-2}(\alpha)$ of the generalized degree of the $(d-2)$-face $\alpha$ over CQNM realizations. The solution of this equation will provide the evolution of the average of the generalized degree of a $(d-2)-$ face $\alpha$ with energy $\epsilon_{\alpha}$ at time $t$, given that the face $\alpha$ is added to   the CQNM at time $t_{\alpha}$, i.e. the solution will specify the function $\hat{k}_{d,d-2}(\alpha)=\hat{k}_{d,d-2}(t,t_{\alpha})$. The mean-field  equation is given by \bea
\frac{d\hat{k}_{d,d-2}({\alpha})}{dt}=\frac{e^{-\beta (\epsilon_{\alpha}-\mu_{d,d-2})}}{t},
\label{Sex}
\eea
with solution $\hat{k}_{d,d-2}(\alpha)=\hat{k}_{d,d-2}(t,t_{\alpha})$ and initial condition $\hat{k}_{d,d-2}(t_{\alpha},t_{\alpha})=1$.
It follows that $k_{d,d-2}(t,t_{\alpha})$ evolves in time as 
\bea
\hat{k}_{d,d-2}(t,t_{\alpha})=e^{-\beta (\epsilon_{\alpha}-\mu_{d,d-2})}\ln\left(\frac{t}{t_{\alpha}}\right)+1.
\eea
The average $\Avg{k_{d,d-2}-1|\epsilon}$ of the generalized degree minus one of $(d-2)-$faces with energy $\epsilon$ is given by 
\bea
\Avg{[k_{d,d-2}-1]|\epsilon}=\Avg{[\hat{k}_{d,d-2}-1]|\epsilon}=\frac{1}{t}\int_{1}^t dt_{\alpha}e^{-\beta (\epsilon-\mu_{d,d-2})}\ln\left(\frac{t}{t_{\alpha}}\right)=e^{-\beta (\epsilon-\mu_{d,d-2})}+{\cal O}(\ln t/t).\nonumber
\eea
Therefore we obtain also in the mean field approximation, that the average generalized degree of $(d-2)-$faces with energy $\epsilon$, for $t\gg1$ satisfies, 
\bea
\Avg{[k_{d,d-2}-1]|\epsilon}=e^{\beta(\epsilon_{\alpha}-\mu_{d,d-2})}=n_{Z}(\epsilon,\mu_{d,d-2}),
\eea
where $n_{Z}(\epsilon,\mu_{d,d-2})$ is proportional to the  Boltzman distribution at temperature $T=1/\beta$. \\
Finally the mean-field equation for the generalized degrees of $\delta$-faces with $\delta<d-2$, is given by 
\bea
\frac{d\hat{k}_{d,\delta}({\alpha})}{dt}=\frac{e^{-\beta (\epsilon_{\alpha}-\mu_{d,\delta})}[\hat{k}_{d,\delta}({\alpha})+2/(d-\delta-2)]}{t},
\label{Splus}
\eea
where one considers the solution $\hat{k}_{d,\delta}({\alpha})=\hat{k}_{d,\delta}(t,t_{\alpha})$ with  initial condition $\hat{k}_{d,\delta}(t_{\alpha},t_{\alpha})=1$.
Here $\hat{k}_{d,\delta}$ indicates the average over CQNM realizations of the average degree of the $\delta-$face $\alpha$, and $\mu_{d,\delta}$ is the chemical potential defined in Eq. $(\ref{Sselfdelta})$.
The mean-field solution of Eq. $(\ref{Splus})$  is given by 
\bea
\hat{k}_{d,\delta}(t,t_{\alpha})-\frac{2}{d-\delta-2}=A\left(\frac{t}{t_{\alpha}}\right)^{e^{-\beta (\epsilon_{\alpha}-\mu_{d,\delta})}},
\eea
where $A=1+{2}/({d-\delta-2})=(d-\delta)/(d-\delta-2)$.
The average $\Avg{[k_{d,d-2}-1]|\epsilon}$ of the generalized degree minus one of the $(d-2)-$faces with energy $\epsilon$ is given by 
\bea
\Avg{[k_{d,\delta}-1]|\epsilon}=\Avg{[\hat{k}_{d,\delta}-1]|\epsilon}=A\frac{1}{t}\int_1^t dt_{\alpha}\left[\left(\frac{t}{t_{\alpha}}\right)^{e^{-\beta (\epsilon-\mu_{d,\delta})}}-1\right]=A\frac{1}{e^{\beta(\epsilon-\mu_{d,\delta})}-1}+{\cal O}(t^{-1+e^{\beta(\epsilon-\mu_{d,\delta})}}).
\nonumber \eea
Therefore we obtain also in the mean field approximation, that the average generalized degree of $\delta-$faces with energy $\epsilon$, and $\delta<d-2$, for $t\gg1$ satisfies, 
\bea
\Avg{[k_{d,\delta}-1]|\epsilon}=A\frac{1}{e^{\beta(\epsilon-\mu_{d,\delta})}-1}=An_{B}(\epsilon,\mu_{d,\delta}),
\eea
where $n_{B}(\epsilon,\mu_{d,\delta})$ is proportional to the  Bose-Einstein distribution at temperature $T=1/\beta$.\\
As a final remark we note  that the mean-field Eqs. $(\ref{Sminus}),(\ref{Sex}),(\ref{Splus})$ can be written by the single equation
\bea
\frac{dy}{dt}=[{a_i}+(1-|a_i|)]e^{-\beta (\epsilon-\mu)}\frac{y^{|a_i|}}{t}
\label{Sgen}
\eea 
with $a_i=-1,0,1$.\\
In fact Eq. $(\ref{Sminus})$ is equal  to Eq. $(\ref{Sgen})$ where $a_i=-1$ and $y=\hat{\xi}_{\alpha}$.
Eq. $(\ref{Sex})$ is equal to Eq. $(\ref{Sgen})$ with $a_i=0$ and $y=\hat{k}_{d,d-1}$.
Finally Eq. $(\ref{Splus})$ is equal to Eq. $(\ref{Sex})$ with $a_i=1$ and $y=\hat{k}_{d,\delta}+2/(d-\delta-2)$.

\end{document}